\documentclass[12pt]{article}

\newcommand{\be}{\begin{equation}}
\newcommand{\ee}{\end{equation}}
\newcommand{\bi}[1]{\vspace{-3mm} \bibitem{#1}}
\usepackage{epsfig,amsmath,amssymb,graphics,graphicx}

\begin{document}

\begin{center}
{\it Journal of Physics A 41 (2008) 435101 }
\end{center}

\begin{center}
{\Large \bf  Fractional Equations of Kicked Systems \\ and Discrete Maps}
\vskip 5 mm

{\large \bf Vasily E. Tarasov$^{1,2}$, and George M. Zaslavsky$^{1,3}$ }\\

\vskip 3mm

{\it $1)$ Courant Institute of Mathematical Sciences, New York University \\
251 Mercer St., New York, NY 10012, USA }\\ 
{\it $2)$ Skobeltsyn Institute of Nuclear Physics, \\
Moscow State University, Moscow 119991, Russia } \\
{\it $3)$ Department of Physics, New York University, \\
2-4 Washington Place, New York, NY 10003, USA } 

\vskip 11 mm

\begin{abstract}
Starting from kicked equations of motion with derivatives of non-integer orders,
we obtain "fractional" discrete maps.
These maps are generalizations of well-known
universal, standard, dissipative, kicked damped rotator maps.
The main property of the suggested fractional maps is a long-term memory.
The memory effects in the fractional discrete maps mean  
that their present state evolution depends on all past states 
with special forms of weights. 
These forms are represented by combinations of power-law functions.
\end{abstract}

\end{center}

\section{Introduction}

There are a number of distinct areas of physics where
the basic problems can be reduced to the study of simple discrete maps. 
Discrete maps have been used for the study of evolution problems, 
possibly as a substitute of differential equations \cite{Chirikov,Schuster,CE,Zaslavsky1}. 
They lead to a much simpler formalism, which is particularly useful in simulations.
The standard map is one of the most widely studied maps.
In this paper, we consider fractional generalizations of discrete maps 
that can be used to study the evolution described 
by fractional differential equations \cite{SKM,Podlubny,KST}. 

The treatment of nonlinear dynamics in terms of discrete maps 
is a very important step in understanding the qualitative
behavior of continuous systems described by differential equations.
Note that the continuous limit of discrete systems with long-range interactions
gives differential equations with derivatives of non-integer orders 
(see for example, \cite{TZ3,JPA2006,JPA2008}).
The derivatives of non-integer orders are a natural generalization of 
the ordinary differentiation of integer order. 
Fractional differentiation with respect to time is
characterized by long-term memory effects that 
correspond to intrinsic dissipative processes 
in the physical systems. 
The application of memory effects to discrete maps means 
that their present state evolution depends on all
past states \cite{Ful,Fick1,Fick2,Giona,Gallas,Stan}.

The mapping $x_{n+1}=f(x_n)$ does not have any memory, as
the value $x_{n+1}$ only depends on $x_n$. 
The introduction of memory means that the discrete value $x_{n+1}$ 
is connected with the previous values 
$x_n$, $x_{n-1}$, . . . , $x_1$. 
Particularly, any system, which is described by a discrete map, 
will have a full memory \cite{Stan}, if each state of the system 
is a simple sum of all previous states:
\be \label{ur1}
x_{n+1} =\sum^n_{k =1} f(x_k),
\ee
where $f(x)$ is a function that defines the discrete map.
In  general, the expression of equation (\ref{ur1}) can tend to infinity.
Note  that  the  full memory exists for functions  that  give  a
finite sum in Eq. (\ref{ur1}).
The full memory is ideal because 
it has the same action upon the next states as all the others in memory.
The map with a long-term memory can be expressed as
\be \label{ur2}
x_{n+1} =\sum^n_{k=1} V_{\alpha}(n,k) f(x_k) ,
\ee
where the weights $V_{\alpha}(n,k)$, and the parameter $\alpha$ 
characterize the non-ideal memory effects.
The  forms of the functions $f(x)$ and $V_a (n,k)$ in equation (\ref{ur2}) 
are obtained by the differential equation. 
Therefore the conditions on $f(x)$ are not discussed. 
Note that linear differential (kicked) equations 
give the linear function $f(x)=x$.
The discrete maps with memory are considered, for example, in the papers
\cite{Ful,Fick1,Fick2,Giona,Gallas,Stan}. 
The important question is a connection of fractional equation 
of motion and the discrete maps with memory.
It is important to derive discrete maps with memory 
from equation of motion with fractional derivatives.

Here, we point out the well-known notions such that 
the full memory and the long-term  memory (LTM). 
For the one-dimensional case, the full memory is defined by equation (\ref{ur1}), 
and the LTM is defined by Eq. (\ref{ur2}).  
In general, the expression of Eq. (\ref{ur1}) can tend to infinity. 
Note that the full memory exists for functions that give a finite sum. 
This is a basic condition on $f(x)$ in Eq. (\ref{ur1}). 
In the paper, we consider the LTM for fractional equations.  
For the one-dimensional case, the LTM is defined by equation (\ref{ur2}). 
The one-dimensional case is considered for simplification. 
In this paper, we  consider a two-dimensional case of variables 
$(x_n,p_n)$. Note that the conditions on the function $f(x)$ 
of the  LTM are not defined {\it a priori}. 
This function is obtained from an equation of motion. 
Therefore the conditions on $f(x)$ are not discussed. 
The properties of the function $f(x)$ are defined by
the kicked  differential equation. 
For example, the  functions $f(x)$, which are derived  
from linear differential (kicked) equations, are linear ($f(x)=x$).

It was shown \cite{ZSE} that perturbed by a periodic force, 
the nonlinear system with a fractional derivative exhibits a new type 
of chaotic motion called the fractional chaotic attractor. 
The fractional discrete maps allow us to study new type of attractors
that are called pseudochaotic \cite{ZSE}. 

In section 2, a brief review of discrete maps is considered 
to fix notation and provide convenient references. 
In section 3, a fractional generalization of the universal map is obtained 
from kicked fractional equations of motion with order $1< \alpha \le 2$.
We prove that the usual universal map is a special case of the 
fractional universal map. 
Some examples of fractional universal map are suggested. 
In section 4, a fractional universal map for the case of $\alpha>2$ is obtained.
In section 5, a fractional kicked damped rotator map is derived.
Finally, a short conclusion is given in section 6.


\section{Universal and standard map}

In this section, a brief review of well-known discrete maps 
is considered to fix notations and provide convenient references. 
For details, see \cite{Chirikov,Schuster,CE,Zaslavsky1}.

Let us consider the equations of motion
\be \label{eq1}
\ddot{x}+K G(x) \sum^{\infty}_{n=0} \delta \Bigl(\frac{t}{T}-n \Bigr)=0
\ee
in which perturbation is a periodic sequence of delta-function-type pulses (kicks)
following with period $T=2\pi / \nu$, $K$ is an amplitude of the pulses, 
and $G(x)$ is a some function.
This equation can be presented in the Hamiltonian form
\be \label{HE1}
\dot{x}=p , \quad
\dot{p}+K G(x) \sum^{\infty}_{n=0} \delta \Bigl(\frac{t}{T}-n \Bigr)=0 .
\ee

It is well-known that these equations can be represented in the form
of discrete map (see for example \cite{Zaslavsky1}).
Between any two kicks there is a free motion
\be \label{sol1}
p=const, \quad x=pt+const . \ee
The solution of the left side of the $n$-th kick
\be \label{not1}
x_n=x(t_n-0)=\lim_{\varepsilon \rightarrow 0+} x(nT-\varepsilon), \quad 
p_n=p(t_n-0)=\lim_{\varepsilon \rightarrow 0+} p(nT-\varepsilon), \quad 
t_n=nT
\ee
is connected with the solution on the right-hand side of the kick
$x(t_n+0)$, $p(t_n+0)$ by Eq. (\ref{HE1}), and 
the condition of continuity $x(t_n+0)=x(t_n-0)$. 
The integration of (\ref{HE1}) over the interval 
$(t_n-\varepsilon,t_n+\varepsilon)$ gives
\[ p(t_n+0)=p(t_n-0)-KT\, G(x_n) . \]
Using notation (\ref{not1}), and the solution (\ref{sol1}), 
we can derive the iteration equations
\[
x_{n+1}=x_n+p_{n+1}T ,
\]
\be \label{E2}
p_{n+1}=p_n-KT\, G(x_n) .
\ee
Equations (\ref{E2}) are called the universal map.

If $G(x)=-x$, then equations (\ref{E2}) give the Anosov-type system
\be
x_{n+1}=x_n+p_{n+1}T , \quad p_{n+1}=KT x_n+p_n .
\ee

For $G(x)=\sin (x)$, equations (\ref{E2}) are
\be
x_{n+1}=x_n+p_{n+1}T , \quad p_{n+1}=p_n-KT \, \sin(x_n) .
\ee
This map is known as the standard or Chirikov-Taylor map \cite{Chirikov}. 


\section{Fractional generalization of the universal map for $1 <\alpha \le 2$}

In this section, a fractional generalization of 
the differential equation (\ref{eq1}) is suggested. 
The discrete map that corresponds to the fractional equation 
of order $1 <\alpha \le 2$ is derived. 
This map can be considered as a generalization of the universal map
for the case $1 <\alpha \le 2$.

Let us consider a fractional generalization of (\ref{eq1}) in the form
\be \label{eq2}
_0D^{\alpha}_t x+K G(x) \sum^{\infty}_{n=0} \delta \Bigl(\frac{t}{T}-n \Bigr)=0, 
\quad (1 <\alpha \le 2) ,
\ee
where $ _0D^{\alpha}_t$ is the Riemann-Liouville fractional derivative
\cite{SKM,Podlubny,KST}, which is defined by
\be \label{RLFD}
_0D^{\alpha}_t x=D^2_t \ _0I^{2-\alpha}_t x=
\frac{1}{\Gamma(2-\alpha)} \frac{d^2}{dt^2} \int^{t}_0 
\frac{x(\tau) d \tau}{(t-\tau)^{\alpha-1}} , \quad (1 <\alpha \le 2) .
\ee
Here we use the notation $D^2_t=d^2/dt^2$, and 
$ _0I^{\alpha}_t$ is a fractional integration \cite{SKM,Podlubny,KST}. \\

\noindent
{\large \bf Proposition 1.}
{\it The fractional differential equation of the kicked system (\ref{eq2})
is equivalent to the discrete map
\be 
p_{n+1}=p_n-KT \, G(x_n) ,
\ee
\be 
x_{n+1}=\frac{T^{\alpha-1}}{\Gamma(\alpha)} \sum^{n}_{k=0} p_{k+1} V_{\alpha}(n-k+1) ,
\quad (1<\alpha\le 2) ,
\ee
where the function $V_{\alpha}(z)$ is defined by}
\be 
V_{\alpha}(z)=z^{\alpha-1}-(z-1)^{\alpha-1} \quad (z \ge 1).
\ee

{\bf Proof.}
Let us define the variable $\xi (t)$ such that
\be \label{xi}
_0^CD^{2-\alpha}_t \xi=x(t),
\ee
where $ _0^CD^{2-\alpha}_t$ is the Caputo fractional derivative
\be \label{CFD}
_0^CD^{2-\alpha}_t \xi=\ _0I^{\alpha-1}_t D^1_t \xi=
\frac{1}{\Gamma(\alpha-1)} \int^{t}_0 
\frac{d \tau}{(t-\tau)^{2-\alpha}} \frac{d \xi(\tau)}{d \tau} ,
\quad (0 \le 2-\alpha<1).
\ee
Using Lemma 2.22 of \cite{KST}, we get
\be \label{Lemma}
_0I^{2-\alpha}_t\ _0^CD^{2-\alpha}_t \xi = \xi(t)-\xi(0) .
\ee
Then
\be \label{xix}
_0D^{\alpha}_t x= D^2_t \ _0I^{2-\alpha}_t x=
D^2_t \ _0I^{2-\alpha}_t \ _0^CD^{2-\alpha}_t \xi=
D^2_t ( \xi(t)-\xi(0) )=D^2_t \xi .
\ee
Substitution of (\ref{xix}) and (\ref{xi}) into Eq. (\ref{eq2}) gives
\be 
D^2_t \xi +K G(\, _0^CD^{2-\alpha}_t \xi ) 
\sum^{\infty}_{n=0} \delta \Bigl(\frac{t}{T}-n \Bigr)=0, 
\quad (1 <\alpha \le 2) .
\ee
This fractional equation can be presented in the Hamiltonian form
\[
\dot{\xi}=\eta ,
\]
\be \label{HE2}
\dot{\eta}+K G(\, _0^CD^{2-\alpha}_t \xi ) 
\sum^{\infty}_{n=0} \delta \Bigl(\frac{t}{T}-n \Bigr)=0 .
\ee

Between any two kicks there is a free motion
\be \label{sol2}
\eta=const, \quad \xi=\eta t+const . \ee
The integration of (\ref{HE2}) over $(t_n+\varepsilon,t_{n+1}-\varepsilon)$ gives
\be \label{z1}
\xi(t_{n+1}-0)=\xi(t_n+0)+\eta_{n+1} T , \quad \eta(t_{n+1}-0)=\eta(t_n+0) .
\ee

The solution of the left side of the $n$-th kick
\be \label{not2}
\xi_n=\xi(t_n-0)=\lim_{\varepsilon \rightarrow 0} \xi(nT-\varepsilon), \quad 
\eta_n=\eta(t_n-0)=\lim_{\varepsilon \rightarrow 0} \eta(nT-\varepsilon), \quad 
t_n=nT,
\ee
is connected with the solution on the right-hand side of the kick
$\xi(t_n+0)$, $\eta(t_n+0)$ by equations (\ref{HE2}), and 
the continuity condition
\be \label{z2} 
\xi(t_n+0)=\xi(t_n-0) . 
\ee
The integration of (\ref{HE2}) over the interval 
$(t_n-\varepsilon,t_n+\varepsilon)$ gives
\be \label{z3}
\eta(t_n+0)=\eta(t_n-0)-KT\, G( \, _0^CD^{2-\alpha}_{t_n} \xi ) . 
\ee
Using notation (\ref{not2}), and the solution (\ref{sol2}), we get
\be
\eta(t_n+0)=\eta(t_{n+1}-0)=\eta_{n+1} .
\ee
Substituting of (\ref{z2}) and (\ref{z3}) into (\ref{z1}), 
we derive the iteration equations
\be
\xi_{n+1}=\xi_n+\eta_{n+1}T ,
\ee
\be \label{F2}
\eta_{n+1}=\eta_n-KT\, G(\, _0^CD^{2-\alpha}_{t_n} \xi ) .
\ee

To derive a map, we should express the fractional derivative 
\be
_0^CD^{2-\alpha}_{t_n} \xi= \frac{1}{\Gamma(\alpha-1)} \int^{t_n}_0 
\frac{d \tau}{(t_n-\tau)^{2-\alpha}} \frac{d \xi(\tau)}{d \tau} 
\ee
through the variables (\ref{not2}). 
Using $\dot{\xi}=\eta$, we have
\be
_0^CD^{2-\alpha}_{t_n} \xi=
\frac{1}{\Gamma(\alpha-1)} \int^{t_n}_0 
\frac{\eta(\tau) d \tau}{(t_n-\tau)^{2-\alpha}} .
\ee
It can be presented as
\be \label{tk}
_0^CD^{2-\alpha}_{t_n} \xi= \frac{1}{\Gamma(\alpha-1)} 
\sum^{n-1}_{k=0} \int^{t_{k+1}}_{t_k} 
\frac{ \eta(\tau) d \tau}{(t_n-\tau)^{2-\alpha}} ,
\ee
where $t_{k+1}=t_k+T=(k+1)T$, $t_k=kT$, and $t_0=0$.

For the interval $(t_k,t_{k+1})$, equations (\ref{sol2}) and (\ref{not2}) give
\be
\eta(\tau)=\eta(t_k+0)=\eta(t_{k+1}-0)=\eta_{k+1}, \quad \tau \in (t_k,t_{k+1}) . 
\ee
Then 
\[ \int^{t_{k+1}}_{t_k} \frac{ \eta(\tau) d \tau}{(t_n-\tau)^{2-\alpha}} =
\eta_{k+1} \int^{t_{k+1}}_{t_k} (t_n-\tau)^{\alpha-2} d\tau = \]
\[  =\eta_{k+1} \int^{t_n-t_k}_{t_n-t_{k+1}} z^{\alpha-2} dz =
\eta_{k+1} \frac{z^{\alpha-1}}{\alpha-1} \Bigl|^{t_n-t_k}_{t_n-t_{k+1}}= \]
\[ =\frac{1}{\alpha-1} \eta_{k+1} 
\Bigl[(t_n-t_k)^{\alpha-1}-(t_n-t_{k+1})^{\alpha-1} \Bigr]= \]

\be \label{tk2}
=\eta_{k+1} \frac{T^{\alpha-1}}{\alpha-1} 
\Bigl[ (n-k)^{\alpha-1}-(n-k-1)^{\alpha-1} \Bigr]. 
\ee
Using $(\alpha-1)\Gamma(\alpha-1)=\Gamma(\alpha)$ and Eq. (\ref{tk2}), 
the fractional derivative (\ref{tk}) can be presented as
\be \label{29}
_0^CD^{2-\alpha}_{t_n} \xi=
\frac{T^{\alpha-1}}{\Gamma(\alpha)} \sum^{n-1}_{k=0} \eta_{k+1} V_{\alpha}(n-k) ,
\quad (1<\alpha\le 2),
\ee
where
\be \label{V}
V_{\alpha}(z)=z^{\alpha-1}-(z-1)^{\alpha-1} \quad (z \ge 1) .
\ee

As a result, Eq. (\ref{F2}) takes the form
\be \label{First}
\xi_{n+1}=\xi_n+\eta_{n+1}T ,
\ee
\be \label{F3}
\eta_{n+1}=\eta_n-KT \,
G\Bigl( \frac{T^{\alpha-1}}{\Gamma(\alpha)} 
\sum^{n-1}_{k=0} \eta_{k+1} V_{\alpha}(n-k) \Bigr) ,
\quad (1<\alpha\le 2).
\ee
These equations define the fractional generalization of the universal map
for variables $(\xi_n, \eta_n)$.


The fractional equation (\ref{eq2}) in the Hamiltonian form can be presented as
\be \label{p1}
\dot{p}+K G(x) \sum^{\infty}_{n=0} \delta \Bigl(\frac{t}{T}-n \Bigr)=0, 
\ee
\be \label{p2}
_0D^{\alpha-1}_t x =p , \quad (1 <\alpha \le 2) ,
\ee
where we use $_0D^{\alpha}_t =D^1_t \, _0D^{\alpha-1}_t$. 

Equations (\ref{xi}) and (\ref{29}) give
\be \label{35}
x_n=\, _0^CD^{2-\alpha}_{t_n} \xi=
\frac{T^{\alpha-1}}{\Gamma(\alpha)} \sum^{n-1}_{k=0} \eta_{k+1} V_{\alpha}(n-k) .
\ee
Using equations (\ref{Lemma}) and (\ref{p2}), we obtain
\be \label{p-eta}
p=\, _0D^{\alpha-1}_t x= D^1_t \ _0I^{2-\alpha}_t x=
D^1_t \ _0I^{2-\alpha}_t \ _0^CD^{2-\alpha}_t \xi=
D^1_t ( \xi(t)-\xi(0) )=D^1_t \xi .
\ee
The definition of $\eta$ in (\ref{HE2}) and Eq. (\ref{p-eta}) give 
\[ p=\dot{\xi}=\eta, \quad p_n=\eta_n . \]
As a result, equations (\ref{F3}), and (\ref{35}) give
\be \label{Main1}
p_{n+1}=p_n-KT \, G(x_n) ,
\ee
\be \label{Main2}
x_{n+1}=\frac{T^{\alpha-1}}{\Gamma(\alpha)} 
\sum^{n}_{k=0} p_{k+1} V_{\alpha}(n-k+1) ,
\quad (1<\alpha\le 2) ,
\ee
where $V_{\alpha}(z)$ is defined in (\ref{V}).

This ends of the proof. $\ \ \ \Box$ \\

Equations (\ref{Main1}) and (\ref{Main2}) 
define the fractional universal map in the phase space $(x_n,p_n)$.
These equations are generalization of the map (\ref{E2}). 

Note that the form of Eqs. (\ref{Main2}) 
is defined by both equations (\ref{p1}) and (\ref{p2}).
Equations (\ref{Main2}) cannot be considered as an iteration representation
of Eq. (\ref{p2}) only. 
If we use the other form of Eq. (\ref{p1}), 
then Eq. (\ref{Main2}) is changed.


Let us consider some examples of 
fractional universal map (\ref{First}), (\ref{F3}). \\


{\bf Example 1.}
Let us prove that the fractional universal map for $\alpha=2$ 
gives the usual universal map. 
Substitution of Eq. (\ref{First}) in the form
\be \label{eq3}
\eta_{k+1}=\frac{1}{T}(\xi_{k+1}-\xi_k) 
\ee
into the iteration equation (\ref{F3}) gives
\be \label{xixi}
\eta_{n+1}=\eta_n-KT \,
G\Bigl( \frac{T^{\alpha-2}}{\Gamma(\alpha)} 
\sum^{n-1}_{k=0} (\xi_{k+1}-\xi_{k}) V_{\alpha}(n-k) \Bigr) .
\ee
Then, the fractional map (\ref{First}), (\ref{F3}) is defined by 
\[ \xi_{n+1}=\xi_n+\eta_{n+1}T , \]
\be \label{FUM2}
\eta_{n+1}=\eta_n-KT \,
G \Bigl( \frac{T^{\alpha-2}}{\Gamma(\alpha)} 
\sum^{n-1}_{k=0} (\xi_{k+1}-\xi_{k}) V_{\alpha}(n-k) \Bigr) , 
\quad (1<\alpha \le 2). 
\ee
For $\alpha=2$, we have $V_{\alpha}(z)=1$, and
\be 
\sum^{n-1}_{k=0} (\xi_{k+1}-\xi_{k}) = \xi_n-\xi_0 .
\ee 
Then equations (\ref{xixi}) gives
\be
\eta_{n+1}=\eta_n-KT \, G \Bigl( \xi_n-\xi_0 \Bigr) .
\ee
As a result, Eqs. (\ref{FUM2}) with $\alpha=2$ give 
the usual universal map (\ref{E2}) for the variables
$x_n=\xi_n-\xi_0$, and $p_n=\eta_n$. \\


{\bf Example 2.}
If $G(x)=-x$, then Eqs. (\ref{First}),  (\ref{F3}) are
\[ \xi_{n+1}=\xi_n+\eta_{n+1}T , \]
\be \label{FAM}
\eta_{n+1}=\eta_n + K 
\frac{T^{\alpha}}{\Gamma(\alpha)} \sum^{n-1}_{k=0} \eta_{k+1} V_{\alpha}(n-k) ,
\quad (1<\alpha\le 2),
\ee
where $V_{\alpha}(z)$ is defined in (\ref{V}).
For the space $(x_n,p_n)$, this generalization can be presented as
\be \label{FAM2} 
p_{n+1}=p_n + K 
\frac{T^{\alpha}}{\Gamma(\alpha)} \sum^{n-1}_{k=0} p_{k+1} V_{\alpha}(n-k+1) ,
\ee
\be \label{FAM3}
x_{n+1}= \frac{T^{\alpha-1}}{\Gamma(\alpha)} 
\sum^{n}_{k=0} p_{k+1} V_{\alpha}(n-k+1) .
\ee
Equations (\ref{FAM2}) and  (\ref{FAM3}) 
define a fractional generalization of the Anosov-type system.  \\

{\bf Example 3.}
If $G(x)=\sin (x)$, equations (\ref{First}),  (\ref{F3}) give
\[ 
\xi_{n+1}=\xi_n+\eta_{n+1}T ,
\]
\be \label{F4}
\eta_{n+1}=\eta_n-KT\, 
\sin \Bigl( \frac{T^{\alpha-1}}{\Gamma(\alpha)} 
\sum^{n-1}_{k=0} \eta_{k+1} V_{\alpha}(n-k) \Bigr) , \quad (1<\alpha \le 2) . 
\ee
This map can be considered as a fractional generalization of the standard map. 
The other possible form of Eq. (\ref{F4}) is 
\[
\xi_{n+1}=\xi_n+\eta_{n+1} ,
\]
\be \label{FSM}
\eta_{n+1}=\eta_n-K \,
\sin \Bigl( \frac{1}{\Gamma(\alpha)} 
\sum^{n}_{k=1} (\xi_{k}-\xi_{k-1}) \Bigl[(n-k)^{\alpha-1}-(n-k-1)^{\alpha-1} \Bigr] \Bigr) ,
\ee
where we use Eqs. (\ref{eq3}), (\ref{V}), $T=1$, and $1<\alpha \le 2$.
For $(x_n,p_n)$, equations (\ref{Main1}), (\ref{Main2}) with $G(x)=\sin (x)$ give
\be 
p_{n+1}=p_n-KT \, \sin ( x_n ) ,
\ee
\be
x_{n+1}= \frac{T^{\alpha-1}}{\Gamma(\alpha)} 
\sum^{n}_{k=0} p_{k+1} V_{\alpha}(n-k+1), \quad (1<\alpha\le 2).
\ee
These equations define a fractional standard map on the phase space, 
which can be called the fractional Chirikov-Taylor map. \\

{\bf Example 4.}
The fractional generalization of the dissipative standard map \cite{DM1,DM2} 
can be defined by 
\be 
\xi_{n+1}=\xi_n+\eta_{n+1} ,
\ee
\be  
\eta_{n+1}=b\eta_n-K \,
\sin \Bigl( \frac{1}{\Gamma(\alpha)} 
\sum^{n}_{k=1} (\xi_{k}-\xi_{k-1}) 
\Bigl[(n-k)^{\alpha-1}-(n-k-1)^{\alpha-1}\Bigr] \Bigr) . 
\ee
For $b=1$, we get the fractional standard map (\ref{FSM}). \\

\section{Fractional universal map for $\alpha>2$}

In this section, a fractional differential equation (\ref{eq2}) 
is used for $\alpha>2$. 
The discrete maps that correspond to the fractional equations 
are derived. 
These maps can be considered as a generalization of the universal map
for the case $\alpha > 2$, i.e., 
the fractional universal map (\ref{First}), (\ref{F3}) 
can be generalized from $1<\alpha\le 2$ to $\alpha>2$. 

Let us consider the fractional equation
\be \label{eq2b}
_0D^{\alpha}_t x+K G(x) \sum^{\infty}_{n=0} \delta \Bigl(\frac{t}{T}-n \Bigr)=0, 
\quad (m-1 <\alpha \le m) ,
\ee
where $ _0D^{\alpha}_t$ is the Riemann-Liouville fractional derivative
of order $\alpha$, $m-1 <\alpha \le m$, which is defied \cite{SKM,Podlubny,KST} by
\be 
_0D^{\alpha}_t x=D^m_t \ _0I^{m-\alpha}_t x=
\frac{1}{\Gamma(m-\alpha)} \frac{d^m}{dt^m} \int^{t}_0 
\frac{x(\tau) d \tau}{(t-\tau)^{\alpha-1}} , \quad (m-1 <\alpha \le m) .
\ee
Here we use the notation $D^m_t=d^m/dt^m$, and 
$ _0I^{m-\alpha}_t$ is a fractional integration \cite{SKM,Podlubny,KST}. \\

\noindent
{\large \bf Proposition 2.}
{\it The fractional differential equation of the kicked system (\ref{eq2b})
is equivalent to the discrete map
\be 
x_n= \frac{1}{\Gamma(\alpha-m+1)} \sum^{n-1}_{k=0} \Bigl(
\sum^{m-3}_{l=0}  \frac{T^{\alpha+l-m+1}}{l!} p^{l+1}_n V^{m,l}_{\alpha}(n-k) +
\frac{T^{\alpha-1}}{(m-2)!} p^{m-1}_{n+1} \, V^{m,m-2}_{\alpha}(n-k) \Bigr) ,
\ee
\be 
p^s_{n+1} =p^s_n +\sum^{m-s-2}_{l=1}  \frac{T^l}{l!} p^{s+l}_n
+ \frac{T^{m-s-1}}{(m-s-1)!} p^{m-1}_{n+1} , \quad (s=1,...,m-2) ,
\ee
\be 
p^{m-1}_{n+1}=p^{m-1}_n-KT\, G(x_n) , \quad (m-1<\alpha \le m) ,
\ee
where the functions $V^{m,l}_{\alpha}(z)$ are defined by
\be 
V^{m,l}_{\alpha}(z) = \int^z_{z-1} (z-y)^l y^{\alpha-m} dy =
\int^1_0 y^l (z-y)^{\alpha-m} dy , \quad l=1,...,m-2 ,
\ee
and $1 \le l \le m-1 < \alpha \le m$, where $l,m \in \mathbb{N}$. } \\

{\bf Proof.}
Let us consider equation (\ref{eq2}) with $m-1<\alpha\le m$.
Using $\xi=\xi(t)$, such that 
\be \label{b0}
 _0^CD^{m-\alpha}_t \xi =x(t), \quad (m-1< \alpha \le m) , \ee
we obtain 
\[ _0D^{\alpha}_t x=\, _0D^{\alpha}_t \, _0^CD^{m-\alpha}_t \xi =
D^m_t \, _0I^{m-\alpha}_t \, _0^CD^{m-\alpha}_t \xi =D^m_t \xi  , \quad (0<m-\alpha<1). \]
Equation  (\ref{eq2}) can be presented as
\be \label{a16}
D^m_t \xi +K G(\, _0^CD^{m-\alpha}_t \xi ) 
\sum^{\infty}_{n=0} \delta \Bigl(\frac{t}{T}-n \Bigr)=0, 
\quad (m-1 <\alpha < m) .
\ee
Let us define
\be \label{etas}
\eta^s(t) =D^s_t \xi (t)  , \quad (s=0,...,m-1) .
\ee
Then the Hamiltonian form of Eq. (\ref{a16}) is
\[
\dot{\eta}^s =\eta^{s+1}  , \quad (s=0,...,m-2) ,
\]
\be
\dot{\eta}^{m-1} + KT\, G(\, _0^CD^{m-\alpha}_{t} \eta^0 )
\sum^{\infty}_{n=0} \delta \Bigl(\frac{t}{T}-n \Bigr)=0 .
\ee

For $t\in (t_n+0,t_{n+1}-0)$, equation (\ref{a16}) is
\[ D^m_t \xi=0 \]
and the solution can be presented as
\be \label{b1}
\xi(t) =\sum^{m-1}_{l=0} C_l (t-t_n)^l , \quad (m \ge 3)
\ee
Substitution of (\ref{b1}) into (\ref{etas}) gives
\be \label{b3}
\eta^s (t) =\sum^{m-1}_{l=s} C_l l(l-1)...(l-s+1)(t-t_n)^{l-s} .
\ee
For $t=t_n$, we have
\[ \eta^s (t_n+0) =  C_s s!  , \]
Then
\be \label{63}
C_l=\frac{1}{l!} \eta^s (t_n+0) 
\ee
Using Eq. (\ref{63}), the relations
\[ \eta^s (t_n+0) = \eta^s (t_n-0) =\eta^s_n , \quad (s=0,...,m-2), \]
and
\[ \eta^{m-1}(t_n+0)=\eta^{m-1}(t_{n+1}-0)=\eta^{m-1}_{n+1} , \]
we present equation (\ref{b3}) in the form
\be \label{b4}
\eta^s (t) =\sum^{m-2}_{l=s}  \frac{1}{(l-s)!} \eta^l_n (t-t_n)^{l-s} +
\frac{1}{(m-s-1)!} \eta^{m-1}_{n+1}(t-t_n)^{m-s-1} ,  \quad (s=0,...,m-2) .
\ee
Here, we use
\be
\frac{l(l-1)...(l-s+1)}{l!} =\frac{1}{(l-s)!} .
\ee
Equation (\ref{b4}) can be rewritten as
\be \label{68}
\eta^s (t) =
\sum^{m-s-2}_{l=0}  \frac{1}{l!} \eta^{s+l}_n (t-t_n)^{l} 
+ \frac{1}{(m-s-1)!} \eta^{m-1}_{n+1}(t-t_n)^{m-s-1} , \quad (s=0,...,m-2) .
\ee
For $t=t_{n+1}$, this equation gives
\be
\eta^s_{n+1} =\sum^{m-s-2}_{l=0}  \frac{1}{l!} \eta^{s+l}_n T^l
+ \frac{1}{(m-s-1)!} \eta^{m-1}_{n+1} T^{m-s-1} , \quad (s=0,...,m-2) .
\ee
As a result, the iteration equations are
\be \label{b-iter1}
\eta^s_{n+1} =\sum^{m-s-2}_{l=0}  \frac{1}{l!} \eta^{s+l}_n T^l
+ \frac{1}{(m-s-1)!} \eta^{m-1}_{n+1} T^{m-s-1} , \quad (s=0,...,m-2) ,
\ee
\be \label{b-iter2}
\eta^{m-1}_{n+1}=\eta^{m-1}_n-KT\, G(\, _0^CD^{m-\alpha}_{t_n} \eta^0 ) .
\ee

The Caputo fractional derivative 
\[  _0^CD^{m-\alpha}_{t_n} \eta^0 = \, _0I^{\alpha-m+1}_{t_n} D^1_t \eta^0 
 = \, _0I^{\alpha-m+1}_{t_n} \eta^1 \]
can be presented as
\be \label{xn}
x_n= \, _0^CD^{m-\alpha}_{t_n} \xi= \frac{1}{\Gamma(\alpha-m+1)} 
\sum^{n-1}_{k=0} \int^{t_{k+1}}_{t_k} 
\frac{ \eta^1(\tau) d \tau}{(t_n-\tau)^{m-\alpha}} ,
\ee
where $t_k=kT$. 

For $t \in (t_k,t_{k+1})$, equation (\ref{68}) is
\be \label{eta1}
\eta^1 (t) =\sum^{m-3}_{l=0}  \frac{1}{l!} \eta^{l+1}_n ( t-t_k)^{l} 
+ \frac{1}{(m-2)!}\eta^{m-1}_{n+1}(t-t_k)^{m-2} .
\ee
Substituting of (\ref{eta1}) into (\ref{xn}), and using
\[ \int^{t_{k+1}}_{t_k} \frac{(t-t_k)^l}{(t_n-t)^{m-\alpha}} dt=
T^{\alpha-m+l+1} \int^{k+1}_k \frac{(\tau-k)^l}{(n-\tau)^{m-\alpha}} d\tau , \]
we obtain
\be
\int^{t_{k+1}}_{t_k} \frac{ \eta^1(t) d t}{(t_n-t)^{m-\alpha}} =
\sum^{m-3}_{l=0}  \frac{T^{\alpha+l-m+1}}{l!} \eta^{l+1}_n V^{m,l}_{\alpha}(n-k) 
+ \frac{T^{\alpha-1}}{(m-2)!} \eta^{m-1}_{n+1} V^{m,m-2}_{\alpha}(n-k)  ,
\ee
where
\be \label{Vml}
V^{m,l}_{\alpha}(n-k) =
\int^{(k+1)}_{k} \frac{(z-k)^l}{(n-z)^{m-\alpha}} d z ,
\quad l=1,...,m-2 .
\ee
These functions can be defined by
\be \label{Vml2}
V^{m,l}_{\alpha}(z) = \int^z_{z-1} (z-y)^l y^{\alpha-m} dy =
\int^1_0 y^l (z-y)^{\alpha-m} dy , \quad l=1,...,m-2 ,
\ee
where $1 \le l \le m-1 < \alpha \le m$, $l,m \in \mathbb{N}$. 

Let us use equation (\ref{b0}) in the form
\[ x_n= \, _0^CD^{m-\alpha}_{t_n} \xi , \quad p^s_n=\eta^s_n , \quad (s=1,...,m-1). \]
As a result, we obtain 
\be \label{b-iter3}
x_n= \frac{1}{\Gamma(\alpha-m+1)} \sum^{n-1}_{k=0} \Bigl(
\sum^{m-3}_{l=0}  \frac{T^{\alpha+l-m+1}}{l!} p^{l+1}_n V^{m,l}_{\alpha}(n-k) +
\frac{T^{\alpha-1}}{(m-2)!} p^{m-1}_{n+1} \, V^{m,m-2}_{\alpha}(n-k) \Bigr) ,
\ee
\be \label{b-iter4}
p^s_{n+1} =p^s_n +\sum^{m-s-2}_{l=1}  \frac{T^l}{l!} p^{s+l}_n
+ \frac{T^{m-s-1}}{(m-s-1)!} p^{m-1}_{n+1} , \quad (s=1,...,m-2) ,
\ee
\be \label{b-iter5}
p^{m-1}_{n+1}=p^{m-1}_n-KT\, G(x_n) , \quad (m-1<\alpha \le m).
\ee
This ends of the proof. $\ \ \ \Box$ \\

{\bf Remark 1.}
Equations (\ref{b-iter3})-(\ref{b-iter5}) define a fractional generalization
of the universal map for $\alpha>2$.
For $G(x_n)=x_n$, equations (\ref{b-iter3})-(\ref{b-iter5}) 
define fractional Anosov-type system with $\alpha>2$. 
For $G(x_n)=\sin x_n$, we have the fractional standard map for $\alpha>2$. \\

{\bf Remark 2.}
The functions that are defined by integrals (\ref{Vml2}) 
can be expressed in elementary functions. 
For example, equation(\ref{Vml}) with $l=0$ gives
\be 
V^{m,0}_{\alpha}(n-k)=\frac{1}{\alpha-m+1}
\Bigl[ (n-k)^{\alpha-m+1}-(n-k-1)^{\alpha-m+1}  \Bigr] .
\ee
This equation for $m=2$ can be presented as
\[ V^{2,0}_{\alpha}(n-k)=\frac{1}{\alpha-1} V_{\alpha} (n-k) . \]
For $l=1$, Eq. (\ref{Vml}) gives
\be
V^{m,1}_{\alpha}(n-k) =\frac{1}{(\alpha-m+1)(\alpha-m+2)}
\Bigl[ (n-k)^{\alpha-m+2}-(n-k-1)^{\alpha-m+1}(n-k+\alpha-m+1)  \Bigr] .
\ee

{\bf Remark 3.}
Note that the function (\ref{Vml2}) can be presented 
through the hypergeometric function (section 2.1.3. of \cite{Bateman})
$F(a,b,c;z)$ by the relation
\be
V^{m,l}_{\alpha}(z) = \frac{1}{(l+1)z^{m-\alpha}} F(m-\alpha,l+1,l+2;z^{-1}) ,
\ee
where 
\[ F(a,b,c;z)= \frac{\Gamma(c)}{\Gamma(b) \Gamma(c-b)}
\int^1_0  \frac{t^{b-1} (1-t)^{c-b-1}}{(1-zt)^a} dt, \]
and we use
\[ \frac{\Gamma(l+1)}{\Gamma(l+2)}=\frac{l!}{(l+1)!}=\frac{1}{l+1} . \]


\section{Fractional kicked damped rotator (FKDR) map}

In this section, a fractional generalization of 
the differential equation for a kicked damped rotator is suggested. 
The discrete map that corresponds to the fractional 
differential equation is derived. 

Let us consider a kicked damped rotator \cite{Schuster}. 
The equation of motion for this rotator is
\be \label{dr}
\ddot{x}+ q \dot{x}=K G(x) \sum^{\infty}_{n=0} \delta \Bigl(t -n T \Bigr) .
\ee
It is well known that this equation gives \cite{Schuster} 
the two-dimensional map
\be \label{dr-map1}
y_{n+1}=e^{-qT} [ y_n+K G(x_n)] ,
\ee
\be \label{dr-map2}
x_{n+1}=x_n+\frac{1-e^{-qT}}{q}  [ y_n+K G(x_n)] .
\ee

Let us consider the fractional generalization of equation (\ref{dr}) in the form 
\be \label{fdr}
_0D^{\alpha}_t x - q \, _0D^{\beta}_t x =
K G(x) \sum^{\infty}_{n=0} \delta (t -n T ) ,
\ee
where
\[ q\in \mathbb{R}, \quad 1< \alpha\le 2 , \quad \beta=\alpha-1 , \]
and $ _0D^{\alpha}_t$ is the Riemann-Liouville fractional derivative
\cite{SKM,Podlubny,KST} defined by (\ref{RLFD}). 
Note that we use the minus on the left-hand side of Eq. (\ref{fdr}), 
where $q$ can have positive and negative values. \\

\noindent
{\large \bf Proposition 3.}
{\it The fractional differential equation of the kicked system (\ref{fdr})
is equivalent to the discrete map
\be 
p_{n+1}=e^{qT} \Bigl[ p_n+K G ( x_n ) \Bigr] ,
\ee
\be 
x_{n+1}=\frac{T^{\alpha-1}}{\Gamma(\alpha-1)} 
\sum^{n}_{k=0} p_{k+1} W_{\alpha}(qT,k-n-1) ,
\ee
where the functions $W_{\alpha}(a,b)$ are defined by 
\be 
W_{\alpha}(a,b)=a^{1-\alpha}  e^{a(b+1)} 
\Bigl[ \Gamma(\alpha-1, ab ) - \Gamma(\alpha-1, a(b+1) ) \Bigr] .
\ee
and $\Gamma(a,b)$ is the incomplete Gamma function:}
\be
\Gamma(a,b)=\int^{\infty}_b y^{a-1} e^{-y} dy  .
\ee

{\bf Proof.}
Let us define the variable $\xi (t)$ such that
\be \label{xi-2}
_0^CD^{2-\alpha}_t \xi=x(t),
\ee
where $ _0^CD^{2-\alpha}_t$ is the Caputo fractional derivative (\ref{CFD}). 
Using 
\be
_0I^{2-\alpha}_t\ _0^CD^{2-\alpha}_t \xi = \xi(t)-\xi(0) , \quad (0\le 2-\alpha <1) ,
\ee
we obtain
\be \label{xix-2}
_0D^{\alpha}_t x= D^2_t \ _0I^{2-\alpha}_t x=
D^2_t \ _0I^{2-\alpha}_t \ _0^CD^{2-\alpha}_t \xi=
D^2_t ( \xi(t)-\xi(0) )=D^2_t \xi ,
\ee
and 
\be \label{xix-3}
_0D^{\beta}_t x= D^1_t \ _0I^{1-\beta}_t x=
D^1_t \ _0I^{2-\alpha}_t x=
D^1_t \ _0I^{2-\alpha}_t \ _0^CD^{2-\alpha}_t \xi=
D^1_t ( \xi(t)-\xi(0) )=D^1_t \xi .
\ee
Substitution of (\ref{xix-2}), (\ref{xix-3}) and (\ref{xi-2}) into Eq. (\ref{fdr}) gives
\be \label{eq7}
D^2_t \xi - q D^1_t \xi=
K G(\, _0^CD^{2-\alpha}_t \xi ) \sum^{\infty}_{n=0} \delta (t-nT), 
\quad (1 <\alpha \le 2) .
\ee

The fractional equation (\ref{eq7}) can be presented in the Hamiltonian form
\[
\dot{\xi}=\eta ,
\]
\be \label{HE3}
\dot{\eta}-q \eta=
K G(\, _0^CD^{2-\alpha}_t \xi ) \sum^{\infty}_{n=0} \delta (t-nT), 
\quad (1 <\alpha < 2, \quad q \in \mathbb{R}) .
\ee

Between any two kicks 
\be \label{67}
\dot{\eta}-q \eta=0 . 
\ee
For $t\in(t_n+0,t_{n+1}-0)$, the solution of equation (\ref{67}) is
\be \label{eq4}
\eta(t_{n+1}-0)=\eta(t_n+0) e^{qT} .
\ee
Let us use the notations 
\be \label{not3}
\xi_n=x(t_n-0)=\lim_{\varepsilon \rightarrow 0} \xi(nT-\varepsilon), \quad 
\eta_n=\eta(t_n-0)=\lim_{\varepsilon \rightarrow 0} \eta(nT-\varepsilon), \quad 
t_n=nT.
\ee

For $t\in (t_n-\varepsilon,t_{n+1}-\varepsilon)$, 
the general solution of (\ref{HE3}) is
\be
\eta(t)=\eta_n e^{q(t-t_n)}+ 
K \sum^{\infty}_{m=0} G(\, _0^CD^{2-\alpha}_{t_m} \xi )  
\int^t_{t_n-\varepsilon} d \tau e^{q(t-\tau)} \delta (\tau-mT) .
\ee
Then
\be \label{eq5}
\eta_{n+1}= e^{q T} \Bigl[ \eta_n+ K G(\, _0^CD^{2-\alpha}_{t_n} \xi ) \Bigr] .
\ee
Using (\ref{eq5}), the integration of the first equation of (\ref{HE3}) gives
\be \label{eq6}
\xi_{n+1}=\xi_n - \frac{1-e^{q T}}{q} 
\Bigl[ \eta_n+ K G(\, _0^CD^{2-\alpha}_{t_n} \xi ) \Bigr] .
\ee

Let us consider the Caputo fractional derivative from Eqs. (\ref{eq5}), (\ref{eq6}),
which is defined by
\be
_0^CD^{2-\alpha}_{t_n} \xi= _0I^{\alpha-1}_t D^1_t \xi=
\frac{1}{\Gamma(\alpha-1)} \int^{t_n}_0 
\frac{d \tau}{(t_n-\tau)^{2-\alpha}} \frac{d \xi(\tau)}{d \tau} , \quad (0\le 2-\alpha <1).
\ee
Using $\dot{\xi}=\eta$, we have
\be
_0^CD^{2-\alpha}_{t_n} \xi= _0I^{\alpha-1}_t \eta
\frac{1}{\Gamma(\alpha-1)} \int^{t_n}_0 
\frac{\eta(\tau) d \tau}{(t_n-\tau)^{\alpha-1}} .
\ee
It can be presented as
\be \label{tk3}
_0^CD^{2-\alpha}_{t_n} \xi= \frac{1}{\Gamma(\alpha-1)} 
\sum^{n-1}_{k=0} \int^{t_{k+1}}_{t_k} 
\frac{ \eta(\tau) d \tau}{(t_n-\tau)^{2-\alpha}} ,
\ee
where $t_{k+1}=t_k+T=(k+1)T$, $t_k=kT$, and $t_0=0$.

For $\tau \in (t_k,t_{k+1})$, equations (\ref{eq4}) and (\ref{not3}) give
\be
\eta(\tau)=\eta(t_k+0) e^{q(\tau-t_k)}=
\eta(t_{k+1}-0) e^{-qT} e^{q(\tau-t_k)}=
\eta_{k+1} e^{q(\tau-t_k-T)}=
\eta_{k+1} e^{q(\tau-t_{k+1})} .
\ee
Then 
\[ 
\int^{t_{k+1}}_{t_k} \frac{ \eta(\tau) d \tau}{(t_n-\tau)^{\alpha-2}} =
\eta_{k+1} \int^{t_{k+1}}_{t_k} 
e^{q(\tau-t_{k+1})} (t_n-\tau)^{\alpha-2} d\tau =
\]

\[ 
=\eta_{k+1} \int^{t_n-t_k}_{t_n-t_{k+1}} 
e^{q(t_n-t_{k+1}-z)} z^{\alpha-2} dz =
\eta_{k+1} e^{q(t_n-t_{k+1})}
\int^{t_n-t_k}_{t_n-t_{k+1}} z^{\alpha-2} e^{-qz} dz =
\]

\be \label{y}
= \eta_{k+1} q^{1-\alpha}  e^{q(n-k-1)T}
\int^{q(t_n-t_k)}_{q(t_n-t_{k+1})} y^{\alpha-2} e^{-y} dy .
\ee

As a result, Eq. (\ref{y}) gives
\be \label{p11}
\int^{t_{k+1}}_{t_k} \frac{ \eta(\tau) d \tau}{(t_n-\tau)^{\alpha-2}} =
\eta_{k+1} q^{1-\alpha}  e^{q(n-k-1)T}
\Bigl[ \Gamma(\alpha-1, q(t_n-t_{k+1}) ) - \Gamma(\alpha-1, q(t_n-t_k) ) \Bigr] .
\ee
Here $\Gamma(a,b)$ is the incomplete Gamma function \cite{Erd}:
\be
\Gamma(a,b)=\int^{\infty}_b y^{a-1} e^{-y} dy , \quad a,b \in \mathbb{C},
\ee
which can be defined by
\be
\Gamma(a,b)=\Gamma(a)-\frac{b^a}{a}\, _1F_1 (1,1+a;-b) ,
\ee
where $ _1F_1$ is the confluent hypergeometric Kummer function \cite{Erd},
\be
_1F_1(a,c;z)=\sum^{\infty}_{k=0} \frac{(a)_k}{(c)_k} \frac{z^k}{k!} .
\ee
Here $(a)_k$ is the Pochhammer symbol
\be
(a)_k=a(a+1)...(a+k-1) , \quad k \in \mathbb{N} .
\ee

Using (\ref{tk3}) and (\ref{p11}), we get
\be \label{tk4}
_0^CD^{2-\alpha}_{t_n} \xi= 
\frac{T^{\alpha-1}}{\Gamma(\alpha-1)} \sum^{n-1}_{k=0} \eta_{k+1} W_{\alpha}(qT,k-n) ,
\quad (1<\alpha \le 2) ,
\ee
where
\be \label{Wa}
W_{\alpha}(a,b)=a^{1-\alpha}  e^{a(b+1)} 
\Bigl[ \Gamma(\alpha-1, ab ) - \Gamma(\alpha-1, a(b+1) ) \Bigr] .
\ee

Substitution of (\ref{tk4}) into (\ref{eq5}) and (\ref{eq6}) gives
\be \label{FDR1}
\eta_{n+1}=e^{qT} \Bigl[ \eta_n+K 
G \Bigl(\frac{T^{\alpha-1}}{\Gamma(\alpha-1)} 
\sum^{n-1}_{k=0} \eta_{k+1} W_{\alpha}(qT,k-n)
 \Bigr) \Bigr] ,
\ee
\be \label{FDR2}
\xi_{n+1}=\xi_n-\frac{1-e^{qT}}{q} \Bigl[\eta_n+
K G \Bigl(\frac{T^{\alpha-1}}{\Gamma(\alpha-1)} 
\sum^{n-1}_{k=0} \eta_{k+1} W_{\alpha}(qT,k-n) \Bigr) \Bigr] .
\ee
Equations (\ref{FDR1}) and (\ref{FDR2}) can be rewritten as
\be \label{FDR3}
\eta_{n+1}=e^{qT} \Bigl[ \eta_n+K 
G \Bigl(\frac{T^{\alpha-1}}{\Gamma(\alpha-1)} 
\sum^{n-1}_{k=0} \eta_{k+1} W_{\alpha}(qT,k-n) \Bigr) \Bigr] ,
\ee
\be \label{FDR4}
\xi_{n+1}=\xi_n+\frac{1-e^{-qT}}{q} \eta_{n+1} .
\ee
These equations can be considered as a fractional generalization of
kicked damped rotator map for $(\xi_n,\eta_n)$.

Equation (\ref{fdr}) can be presented in the Hamiltonian form
\be 
_0D^{\alpha-1}_t x=p, \quad 
\dot{p} - q \, _0D^{\beta}_t x =
K G(x) \sum^{\infty}_{n=0} \delta (t -n T ) .
\ee
For $(x_n,p_n)$, equations (\ref{xi-2}), (\ref{xix-2}), (\ref{xix-3}) give
\[ p_n=\eta_n, \quad x_n=\, _0^CD^{2-\alpha}_t \xi \]
As a result, we have
\be \label{New1}
p_{n+1}=e^{qT} \Bigl[ p_n+K G ( x_n ) \Bigr] ,
\ee
\be \label{New2}
x_{n+1}=\frac{T^{\alpha-1}}{\Gamma(\alpha-1)} 
\sum^{n}_{k=0} p_{k+1} W_{\alpha}(qT,k-n-1) ,
\ee
where $W_{\alpha}(a,b)$ is defined in (\ref{Wa}).
This ends of the proof. $\ \ \ \Box$ \\

These iteration equations (\ref{New1}) and (\ref{New2})
define the fractional kicked damped rotator map (\ref{dr})
in phase space $(x_n,p_n)$.

\section{Conclusion}

There are a number of distinct areas of physics where
the basic problems can be reduced to the study of simple symplectic maps. 
In particular the special case of two-dimensional 
symplectic maps has been extensively studied. 
Under a wide range of circumstances such maps give rise to chaotic behavior. 
The suggested fractional maps can be considered as 
a fractional generalization of the symplectic map that is derived 
from kicked fractional differential equations. 
We can suppose that fractional discrete maps can be connected with
some generalization of the symplectic structure.
Fractional generalization of symplectic structure \cite{Tar}
can be defined by using fractional differential forms.


\section*{Acknowledgments}

This work was supported by the Office of Naval Research, 
Grant No. N00014-02-1-0056, and the NSF Grant No. DMS-0417800.



\end{document}